\documentstyle[prl,aps,twocolumn,psfig,floats]{revtex}

\begin{document}

\title{Negative heat capacities and first order phase transitions in nuclei and
other mesoscopic systems}

\author{ L. G. Moretto, J. B. Elliott, L. Phair, and G. J. Wozniak }

\address{ Nuclear Science Division, Lawrence Berkeley National Laboratory,
Berkeley, CA 94720\\ }

\date{\today}

\maketitle

\begin{abstract}
The origin of predicted and observed anomalies in caloric curves of
nuclei and other mesoscopic systems is investigated.\ \ It is shown
that a straightforward thermodynamical treatment of an evaporating
liquid drop leads to a backbending in the caloric curve and to
negative specific heats in the two phase coexistence region.\ \ The
cause is found not in the generation of additional surface, but in the
progressive reduction of the drop's radius, and surface, with
evaporation.
\end{abstract}


\narrowtext

The thermodynamical equilibrium properties of first order phase
transitions are completely describable in terms of the thermodynamic
state variables associated with the individual separate phases.\ \
This is not the case in continuous phase transitions, where the two
phases become progressively more similar as the critical point is
approached.\ \ For this reason, in contrast with continuous phase
transitions, first order phase transitions are ``trivial,'' and
interesting only in so far as they herald the appearance of a hitherto
unknown or undescribed phase.

Renewed attention to phase transitions has been generated by studies
of models with well defined Hamiltonians with either short range
interactions (e.g. the Ising model
\cite{coniglio,kertesz,wang1,wang2,huller2} or the lattice gas model
\cite{pan1,pan2,campi,muller,richert,carmona,gulminelli,chomaz2}) or
incorporating long range interactions such as gravitation or
electro-magnetic interactions
\cite{richert,hiley,fisher1,bayong,miller,bresme,jund}.\ \ Several of
these studies, microcanonical and canonical, were performed
numerically, thus the results apply directly to finite (mesoscopic)
systems.\ \ Features expected to disappear in the thermodynamic limit,
if such a limit exists, were noticed and were claimed to be essential,
characteristic indicators of phase transitions in mesoscopic systems
\cite{gulminelli,chomaz2,gross1,huller,promberger,doye,wales,gross2,gross3,bondorf,chbihi,leferve}.
For instance, first order phase transitions were associated with
anomalous convex intruders in the entropy versus energy curves,
resulting in backbendings in the caloric curve, and in negative heat
capacities
\cite{huller2,gulminelli,chomaz2,gross1,huller,promberger,doye,wales,gross2,gross3,bondorf,chbihi,leferve,chomaz,dagostino2,schmidt-01}.

It is often claimed that these features appear only in microcanonical
calculations and are thought to become lost or smeared out in the haze
of canonical calculations
\cite{huller2,gross1,huller,promberger,gross2,gross3}.\ \ These
anomalies have been attributed to a variety of causes, the foremost of
which are surface effects (the energetic cost of an interface), and
long range forces \cite{gross1,wales,gross2,gross3}.\ \ Unfortunately
however, the numerical nature of the calculations tends to make the
identification of the causes of negative heat capacities rather
problematic.

In the experimental arena, a very elegant study has permitted the
observation of the melting of clusters of 147 sodium atoms and the
associated negative heat capacity \cite{schmidt-01}.

In the context of nuclear physics, microcanonical models of nuclear
multifragmentation have associated the anomalies of a convex intruder
with the onset of multifragmentation \cite{gross1,gross2,gross3}.\ \
Furthermore, lattice gas models in the isobaric regime have also shown
negative specific heats in the coexistence region, where
multifragmentation also appears \cite{gulminelli,chomaz2}.\ \ The
question of whether the two transitions are related and possibly
coincident with the liquid-vapor transition is still very much open.

Recently, the claim has been made of an empirical observation of these
anomalies, such as negative heat capacities in nuclear systems
\cite{dagostino2}.\ \ These negative heat capacities have been
inferred from the study of fluctuations in multifragmenting nuclear
systems.\ \ Thus there is a great deal of interest in elucidating the
origin of such anomalies in models as well as experiments.

In particular, it would be highly desirable to ground any evidence for
these anomalies, theoretical or otherwise, on thermodynamics itself,
minimally modified to allow for the possible role of surface effects
related to the finiteness of the system.

In this paper we illustrate analytically how effects such as negative
heat capacities can arise within a standard thermodynamic treatment.\
\ As an example, we consider the evaporation of a drop of ordinary
liquid.\ \ Our only concern with mesoscopicity is the explicit
treatment of the surface of the drop.

Nuclear systems have long been associated with liquids, as testified
by the success of the liquid drop model.\ \ In the spirit of the
liquid drop model, the surface energy introduces the simplest (and
dominant!)  correction to the bulk energy, leading to a model of one
percent accuracy in systems as small as 40 nucleons and possibly
smaller.\ \ A similar approach should hold for other kinds of clusters
for which the surface energy may also be the first and dominant
correction to the bulk properties.\ \ Specifically, we will study the
role of surface in generating the anomalies in the caloric curve.

\begin{figure} [ht]
	\centerline{\psfig{file=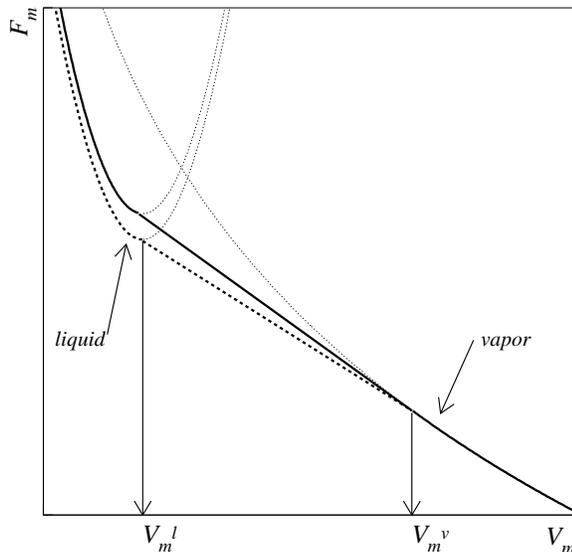,width=7.5cm,angle=0}}
	\caption{A schematic plot of the molar free energy as a
	function of molar volume for a liquid and vapor.\ \ Solid
	(dashed) lines demonstrate the behavior of a droplet (the
	bulk).\ \ Vertical lines show the molar volumes of the liquid
	and vapor at coexistence.\ \ Dotted curves represent
	superheated liquid and supersaturated vapor.}
	\label{fig:free_e}
	\end{figure}

For many years, it has been known that a ``mesoscopic'' system, such
as a tiny drop of liquid with radius $r$ can be readily described in
the pure thermodynamic limit
\cite{rayleigh,tolman,nijmeijer,brodskaya,koga}.\ \ The complete
analogy between the liquid-vapor phase coexistence of a liquid in the
bulk and for a drop can be seen in Fig.~\ref{fig:free_e}, where the
molar free energy $F_m$ at constant $T$ is plotted versus molar volume
$V_m$ for the two phases.\ \ The two free energy branches, for liquid
and for vapor, can be considered completely independent.\ \ No
interaction is assumed between the two phases.\ \ Coexistence becomes
possible in the region of volume $V$ where the overall free energy can
be minimized through the common tangent construction.\ \ The
equilibrium pressure of the saturated vapor is immediately given by
	\begin{equation}
	p = \left. - \frac{\partial F}{\partial V} \right|_{T} .
	\label{pressure-3}
	\end{equation}
No qualitative difference in the picture results by considering a drop
of finite radius $r$.\ \ The only difference is that the overall free
energy of the drop (solid curves) as we shall see below, is higher
than that of the bulk (dashed curves) and the equilibrium vapor
pressure is correspondingly higher.

The state of equilibrium between a liquid and its vapor can be
described in the simplest way by the Clapeyron Equation
	\begin{equation}
	\frac{dp}{dT}=\frac{\Delta H_{m}}{\Delta V_{m}T}:
	\label{clap-eq}
	\end{equation}
where, $p$ and $T$ are the pressure and temperature, $\Delta H_{m}$ is
the molar enthalpy of vaporization and $\Delta V_{m}$ is the
difference of the molar volumes of vapor, $V_{m}^{v}$, and liquid,
$V_{m}^{l}$.

Specialization to the case of a drop of radius $r$ can be achieved by
modifying the enthalpy to account for the surface energy
\cite{moretto}
	\begin{equation}
	\Delta H_{m}=\Delta H_{m}^{0}-c_{s}S_{m}^{l} = \Delta 
	H_{m}^{0}-\frac{3 c_{s} V_{m}^{l}}{r}
	\label{surf-enth}
	\end{equation}
where $\Delta H_{m}^{0}$ is the ``bulk'' molar enthalpy, $S_{m}^{l}$
and $V_{m}^{l}$ are the surface and volume of the drop and $c_{s}$ is
the surface energy coefficient.

Neglecting $V_{m}^{l}$ compared to $V_{m}^{v}$ and considering the
vapor ideal, i.e.\ \ $V_{m}^{v} = T/p$, we can integrate
Eq.~(\ref{clap-eq}), assuming also $\Delta H_{m}$ to be constant.\ \
We obtain 
	\begin{equation}
	p = p_{0}\exp\left( -\frac{\Delta H_{m}^{0}}{T}+\frac{3c_{s}V_{m}^{l}}{rT}\right)
	\label{pressure-1}
	\end{equation}
or
	\begin{equation}
	p = p_{\rm bulk}\exp\left( \frac{3c_{s}V_{m}^{l}}{rT}\right).
	\label{pressure-2}
	\end{equation}
This equation contains \underline{all} the thermodynamical information
necessary to characterize the phase coexistence of the liquid drop of
radius $r$ with its vapor.\ \ The salient feature is the rise of the
vapor pressure with decreasing radius.\ \ Fig.~ref{fig:pt} gives a map
of the function $p' = p'(T',r')$, in terms of the scaled variables
	\begin{equation}
	p' = \frac{p}{p_0} ,~
	T' = \frac{T}{\Delta H_{m}^{0}} ,~
	r' = \frac{\Delta H_{m}^{0}}{3 c_s V_{m}^{l}} r .
	\label{reduced_units}
	\end{equation}
For any given radius $r$, the function $p = p(T,r)$ describes the
equilibrium condition between the drop and its vapor.\ \ In other
words, it is the phase diagram of the drop.\ \ The drop appears here
as a ``phase'' defined by its radius $r$.\ \ A change in radius
implies a change in phase.

	\begin{figure} [ht]
	\centerline{\psfig{file=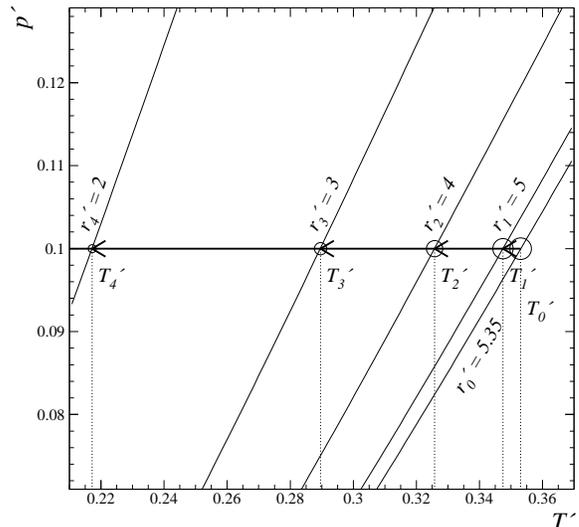,width=7.5cm,angle=0}}
	\caption{The saturated vapor pressure as a function of the
	temperature for different droplet radii.\ \ The size of the
	open circles is proportional to the droplet radius.}
	\label{fig:pt}
	\end{figure}

\begin{figure} [ht]
	\centerline{\psfig{file=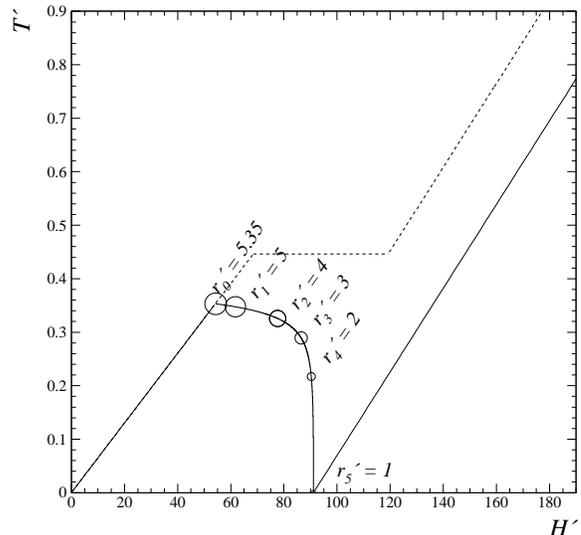,width=7.5cm,angle=0}} 
	\caption{The history dependent caloric curve of an evaporating
	drop at constant pressure.\ \ Dashed lines represent bulk
	behavior, solid line shows the drop's caloric curve.\ \ The
	scaled enthalpy is $H' = H ( \Delta H^{0}_{m} / 3 c_s
	V_{m}^{l} )^3 / ( 4 \pi / V_{m}^{l} )$.}
	\label{fig:th}
	\end{figure}

\begin{figure} [ht]
	\centerline{\psfig{file=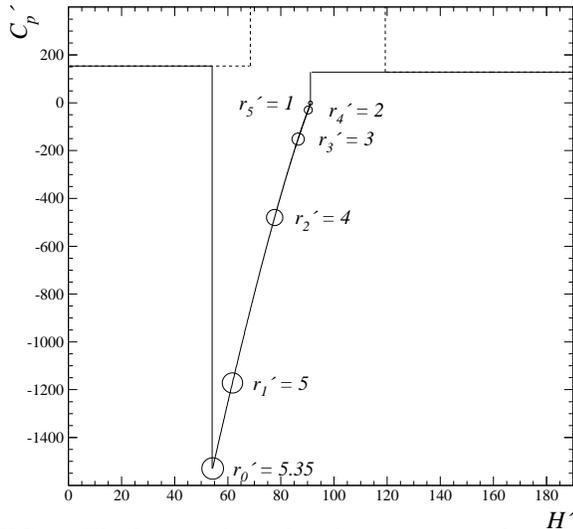,width=7.5cm,angle=0}}
	\caption{The history dependent heat capacity of an evaporating
	drop at constant pressure.\ \ Dashed lines represent bulk
	behavior, solid line shows the drop's caloric curve.\ \ The
	scaled heat capacity is $C_{p}' = C_{p} ( \Delta H^{0}_{m} / 3
	c_s V_{m}^{l} )^3 / ( 4 \pi / \Delta H_{m}^{0} V_{m}^{l} )$.}
	\label{fig:cph}
	\end{figure}

Let us now introduce some history and construct a caloric curve for a
drop of radius $r_{0}$.\ \ In order to operate at constant pressure
$p_{0}$, we enclose the spherical drop in a deformable and expandable
container to which we apply an external pressure $p_{0}$.\ \ As the
drop is heated, and before the vapor can appear, the temperature
increases according to
	\begin{equation}
	\Delta H = C^{l}_{p} \Delta T ,
	\label{enthalpy-1}
	\end{equation}
where $C^{l}_{p}$ is the liquid's heat capacity and is approximately
constant.\ \ When $T$ reaches the value $T_{0}$ at which the vapor
pressure $p(r_{0}) = p_{0}$, the vapor first appears and it expands
against the container.\ \ The heat of vaporization is absorbed at a
rate $H_{m}(r_{0})$.\ \ However, as it evaporates, the drop sees its
radius decreasing from its initial value, here chosen to be $r_{0}' =
5.35$.\ \ At constant temperature the vapor pressure would rise, but,
at constant pressure, as we are now operating, the temperature
\underline{decreases} as shown in Fig.~\ref{fig:pt}, as the system
absorbs its heat of vaporization, so that,
	\begin{eqnarray}
	\Delta H & = & \int_{r}^{r_0} \Delta H_{m}
	\frac{dV}{V_{m}^{l}} \nonumber \\ & = & \frac{4
	\pi}{V_{m}^{l}} \left( \frac{3 c_s V_{m}^{l}}{\Delta
	H^{0}_{m}} \right)^{3} \left[ \frac{1}{3}(r_{0}'^{3} - r'^{3})
	- \frac{1}{2}(r_{0}'^{2} - r'^{2}) \right]
	\label{enth-const-p}
	\end{eqnarray}
and
	\begin{equation}
	T' = T_{0}' \left( \frac{1-\frac{1}{r'}}{1-\frac{1}{r_{0}'}} \right) .
	\label{temp-const-p}
	\end{equation}
After the drop has completely evaporated, the vapor can increase its
temperature according to
	\begin{equation}
	\Delta H = C^{v}_{p} \Delta T
	\label{enthalpy-2}
	\end{equation}
where $C^{v}_{p}$ is the vapor heat capacity at constant pressure.\ \
The resulting caloric curve defined parametrically by
Eq.~(\ref{enth-const-p}) and Eq.~(\ref{temp-const-p}) and shown in
Fig.~\ref{fig:th} is rather interesting.\ \ It has a decreasing branch
associated with the phase transition, along which the heat capacity is
negative!\ \ (See Fig.~\ref{fig:cph}.)

As an aside we note that the scaled radius $r'$ is just the ratio of
the bulk energy to the surface energy.\ \ Thus for a nuclear system
the range shown: $1 \le r_{0}' \le 5.35$, corresponds to a gold
nucleus ($A = 197$) evaporating to a single nucleon.

These rather extraordinary features are wholly due to the interesting
but, in a way, accidental history of the decreasing radius with
increasing evaporation.

Because of the surface energy effect, each drop, of a given radius
$r_0$, is a separate phase in and of itself, different from that
associated with a different radius $r_1$.\ \ At fixed radius $r_0$,
nothing anomalous appears in Fig.~\ref{fig:pt}; the pressure versus
temperature curve, caloric curve and heat capacity are all perfectly
normal.\ \ Anomalies arise when the system drifts from one radius to
another, or from one phase to another.

Typically, experiments
\cite{pochodzalla,schmidt1,hauger2,schmidt2,hauger3} and calculations
\cite{huller2,gulminelli,chomaz2,gross1,huller,promberger,doye,wales,gross2,gross3,bondorf,chbihi,leferve,chomaz,dagostino2}
heat a preassigned system with a certain amount of energy $\Delta E$
or $\Delta H$, and determine the resulting change in entropy $S$ and
temperature $1/T = \partial S / \partial E |_{V}$; from the resulting
caloric curve, phase coexistence diagrams are extracted.\ \ However,
as shown above, the evolution of the system occurring during heating
introduces complications in the construction of a phase diagram from
such a caloric curve.

To avoid this problem an experimentalist or theorist would have to
keep the radius constant or correct for its change while determining
the vapor pressure as a function of the temperature, thereby
eliminating the accidental aspects associated with the evolution of
the system.\ \ The proper representation of all the thermo-physical
properties associated with the coexistence liquid drop-vapor is that
given in Eq.~(\ref{pressure-1}) and Fig.~\ref{fig:pt}.

The results obtained here are firmly grounded on thermodynamics with a
straightforward accounting of finiteness through the surface
correction.\ \ They are exact in the limit in which the liquid drop
model holds, namely, down to nuclei/clusters containing 20 or so
constituents.\ \ They are completely general, as they do not depend on
specific details of the system but rather on its gross properties.\ \
In fact, they should be used as the paragon for lattice gas models and
the like.\ \ In the limit in which these models represent liquid vapor
coexistence, they must reproduce the present results.

Even more importantly, this approach obviates the need for repeating
numerical calculations for each individual system or drop size.\ \ All
that is required is to determine the bulk energy (enthalpy) and the
surface energy coefficient of a give phase once and for all.

It is not clear to us at the moment if the transition studied here and
the anomalies associated with it have a direct counterpart in the
allegedly observed phase transition in nuclear systems
\cite{dagostino2}.\ \ It is however worth repeating that, once the
constraint of constant pressure is enforced, the results described
here are entirely general, as they apply to any small system
undergoing solid-vapor or liquid vapor transitions.

Anomalies in the heat capacities observed in microcanonical
calculations have been attributed to the \underline{increase} in
surface generated as \underline{additional} liquid-vapor interface,
e.g. in the formation of bubbles \cite{gross3}.\ \ In the present
case, however, and possibly generally, this conclusion is not valid.\
\ In a finite system undergoing a liquid-vapor transition there is on
average a \underline{decrease} of surface as the evaporation
proceeds.\ \ Any interior vapor bubble formation is disfavored by a
Boltzmann factor $\exp ( - \Delta V/T )$, compared to the location of
the same vapor on the outside of the drop, whose surface area ends up
actually \underline{decreasing}.\ \ Thus, the resulting anomalies are
indeed surface related, but in a very different way.

Several conclusions and caveats can be drawn from this treatment of
an evaporating liquid drop:

	\begin{enumerate}
	\item Mesoscopic systems can be dealt with within the context
	of standard thermodynamics, minimally modified to include the
	surface.
	\item Anomalous features such as a backbending caloric curve
	and attending negative heat capacities can be made to appear
	by allowing the system to ``evolve'' in parameter space
	(e.g. $r$).
	\item These features appear in a strict thermodynamical
	treatment and thus are \underline{not} specific features of
	microcanonicity.
	\item These features are, in a way, accidental.\ \ They
	reflect the evolution of the system in parameter space (here,
	the radius $r$).\ \ If the radius is kept fixed and the system
	is confined to a ``single'' liquid phase, the phase
	coexistence diagrams are completely ordinary, and no new
	thermodynamics is evident.
	\end{enumerate}

This work was supported by the Director, Office of Energy Research,
Office of High Energy and Nuclear Physics, Nuclear Physics Division of
the US Department of Energy, under Contract No. DE-AC03-76SF00098.

%\newpage

%\vspace{12pt} \noindent\\ Figure 1. \ \ A schematic plot of the molar free
%energy as a function of molar volume for a liquid and vapor.\ \ Solid (dashed)
%lines demonstrate the behavior of a droplet (the bulk).\ \ Vertical lines show
%the molar volumes of the liquid and vapor at coexistence.\ \ Dotted curves
%represent superheated liquid  and supersaturated vapor.

%\vspace{12pt} \noindent\\ Figure 2. \ \ Saturated vapor pressure as a function
%of the temperature for different droplet radii.\ \  The size of the open
%circles is proportional to the droplet radius.

%\vspace{12pt} \noindent\\ Figure 3. \ \ The history dependent caloric curve of
%an evaporating drop at constant pressure.\ \ Dashed lines  represent bulk
%behavior, solid line shows the drop's caloric curve.\ \ The scaled enthalpy is 
%$H' = H ( \Delta H^{0}_{m} / 3 c_s V_{m}^{l} )^3 / ( 4 \pi / V_{m}^{l} )$.

%\vspace{12pt} \noindent\\ Figure 4. \ \ The history dependent heat capacity of
%an evaporating drop at constant pressure.\ \ Dashed lines  represent bulk
%behavior, solid line shows the drop's caloric curve.\ \ The scaled heat
%capacity is  $C_{p}' = C_{p} ( \Delta H^{0}_{m} / 3 c_s V_{m}^{l} )^3 / ( 4 \pi
%/ \Delta H_{m}^{0} V_{m}^{l} )$.

\end{document}